\documentstyle[11pt,epsfig]{article}
\RequirePackage{graphicx}
\RequirePackage{amssymb}
\newcommand{\be}{\begin{equation}}
\newcommand{\ee}{\end{equation}}
\newcommand{\bea}{\begin{eqnarray}}
\newcommand{\eea}{\end{eqnarray}}
\newcommand{\barr}{\begin{eqnarray}}
\newcommand{\earr}{\end{eqnarray}}
\newcommand{\rar}{\rightarrow}
\newcommand{\pdup}{p_\uparrow}
\newcommand{\rpip}{\pi^- + p_\uparrow}

\newcommand{\npionx}{\pi^0 + X}

\newcommand{\pplup}{\pdup + p \rar \pi^+ + X}

\newcommand{\pminp}{\pdup + p \rar \pi^- + X}

\newcommand{\pdupp}{\pdup + p \rar \pi^0 + X}
\newcommand{\pimp}{\pi^- + \pdup \rar \pi^0 + X}

\newcommand{\ppdup}{p + \pdup \rar \pi^0 + X}
\newcommand{\PRD}[1]{{\it Phys.\ Rev.\ D}\ {\bf #1}}

\newcommand{\PLB}[1]{{\it Phys.\ Lett.\ B}\ {\bf #1}}

\newcommand{\xf}{x_{\mathrm F}}
\begin{document}

\begin{center}
{\bfseries Universal threshold for single-spin asymmetries in fixed 
target experiments}
\vskip 3mm
{V.V. Mochalov$^{\dag}$ and A.N. Vasiliev}
\vskip 3mm
{\small {\it Institute of High Energy Physics, Protvino, Russia}

$\dag$ {\it
E-mail: mochalov@mx.ihep.su}}
\end{center}

\begin{abstract}
{The analysis of inclusive $\pi$-meson single spin asymmetry measurements 
was done. It was found that the single spin asymmetry starts to grow up 
at the same value of the $\pi$-meson energy in the center of mass system 
for fixed target experiments at the beam energy range from 13 to 200~GeV.}
\end{abstract}

Polarization experiments give us an unique opportunity to probe the 
nucleon internal structure. While spin averaged cross-sections can be
calculated within acceptable accuracy, current theory of strong 
interactions can not describe large spin asymmetries and polarization. 

Unexpected large values of  single spin asymmetry $A_N$ (SSA)
in inclusive $\pi$-meson production are real challenge
to the theory because perturbative Quantum Chromodynamics 
predicts small asymmetries decreasing with transverse momentum. 
Various models were developed to explain results from E704 (FNAL), 
PROZA-M and FODS (both Protvino) and several BNL experiments. 
Most of the models analyse experimental data in terms of $\xf$ and/or $p_T$.  
To investigate the dependence of SSA on a secondary meson production
angle, the measurements  in the reaction $\pimp$ 
were carried out at the PROZA-M experiment (Protvino) at 40~GeV pion 
beam in the two different kinematic regions:
at Feinman scaling variable $\xf \approx 0$ \cite{protv88} and 
in the polarized target fragmentation region \cite{proza40}.  
It was reported \cite{proza40} that the asymmetry of 
inclusive $\pi^0$ production in the reaction $\pimp$  starts to grow up 
at the same centre of mass energy $E_0^{cms} \approx 1.7$~GeV for the
both kinematic regions. 
A similar behaviour was also found in the reaction $\ppdup$
at 70~GeV \cite{prelim70}.
There is an indication that the asymmeties are zero till some 
threshold value and then start to rise up linearly and in 
principle may saturate at some level.  

In this case we can fit SSA by the function 

\be
A_N = \left\{ \begin{array}{ll} 0 & \textrm {, если $E<E_0$} \\
k \cdot (E-E_0) & \textrm {, если $E \geq E_0$} 
\end{array} \right.
\label{eq:fit}
\ee

\noindent with two parameters -- a threshold energy $E_0$ 
and $k$. Let us mention that  $\pi^0$-mesons were detected 
in rather narrow solid angle. In this case the 
dependence of the asymmetry on energy reflects the dependence
on transverse momentum 
in the central region or on  $\xf$ in the polarized 
particle fragmentation region.
Saturation, if exists, is achieved at large values of
transverse energy or  $\xf$. The error bars in this region are large 
and the last measured points
are not crucial for the results of a fit. In this case we can neglect by the 
saturation effect and use all points for fitting by the function 
(\ref{eq:fit}). 

\begin{figure}[thb] 
\centering
\begin{tabular}{cc} 
\includegraphics[width=0.4\textwidth]
{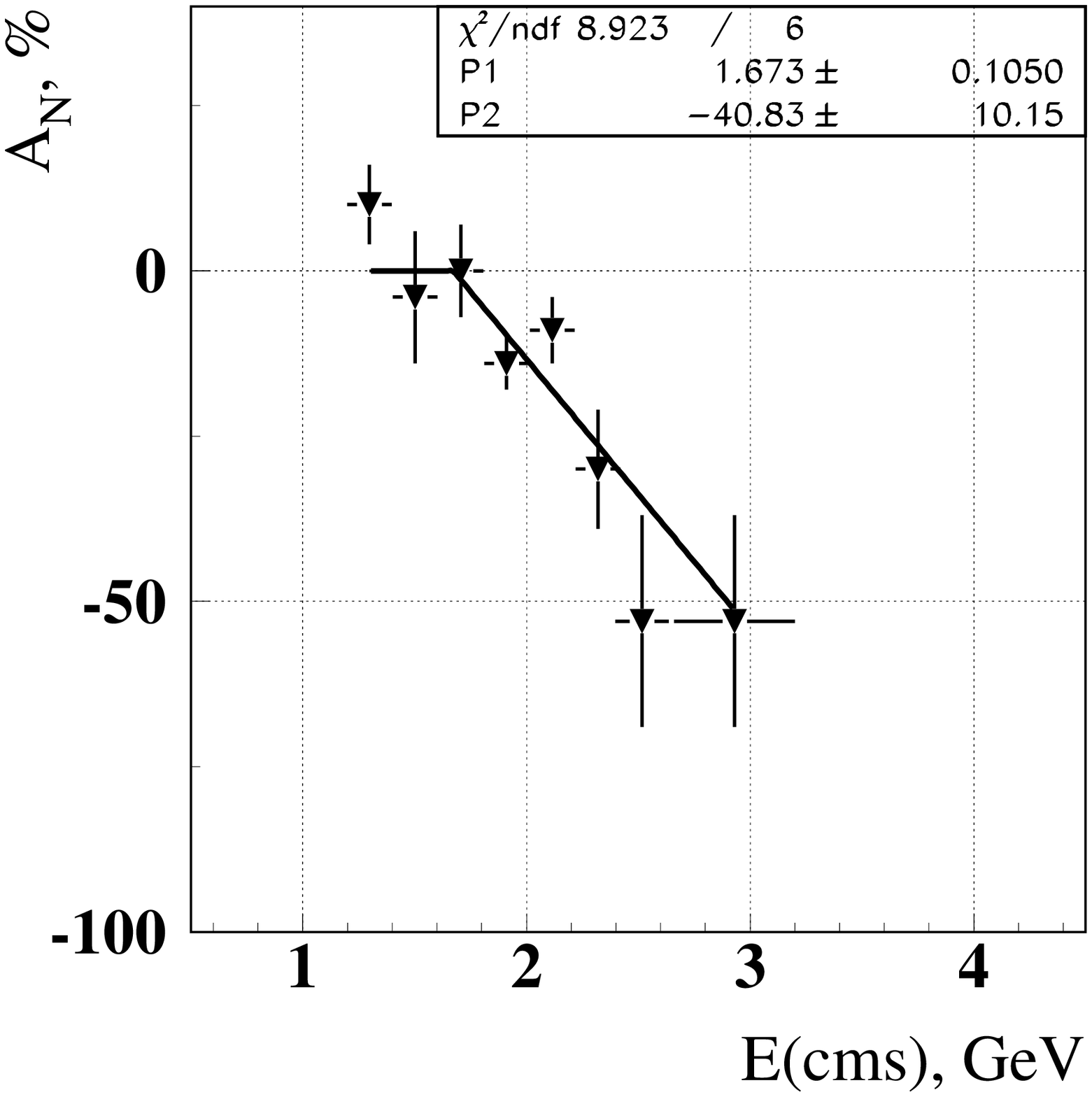} &
\includegraphics[width=0.4\textwidth]
{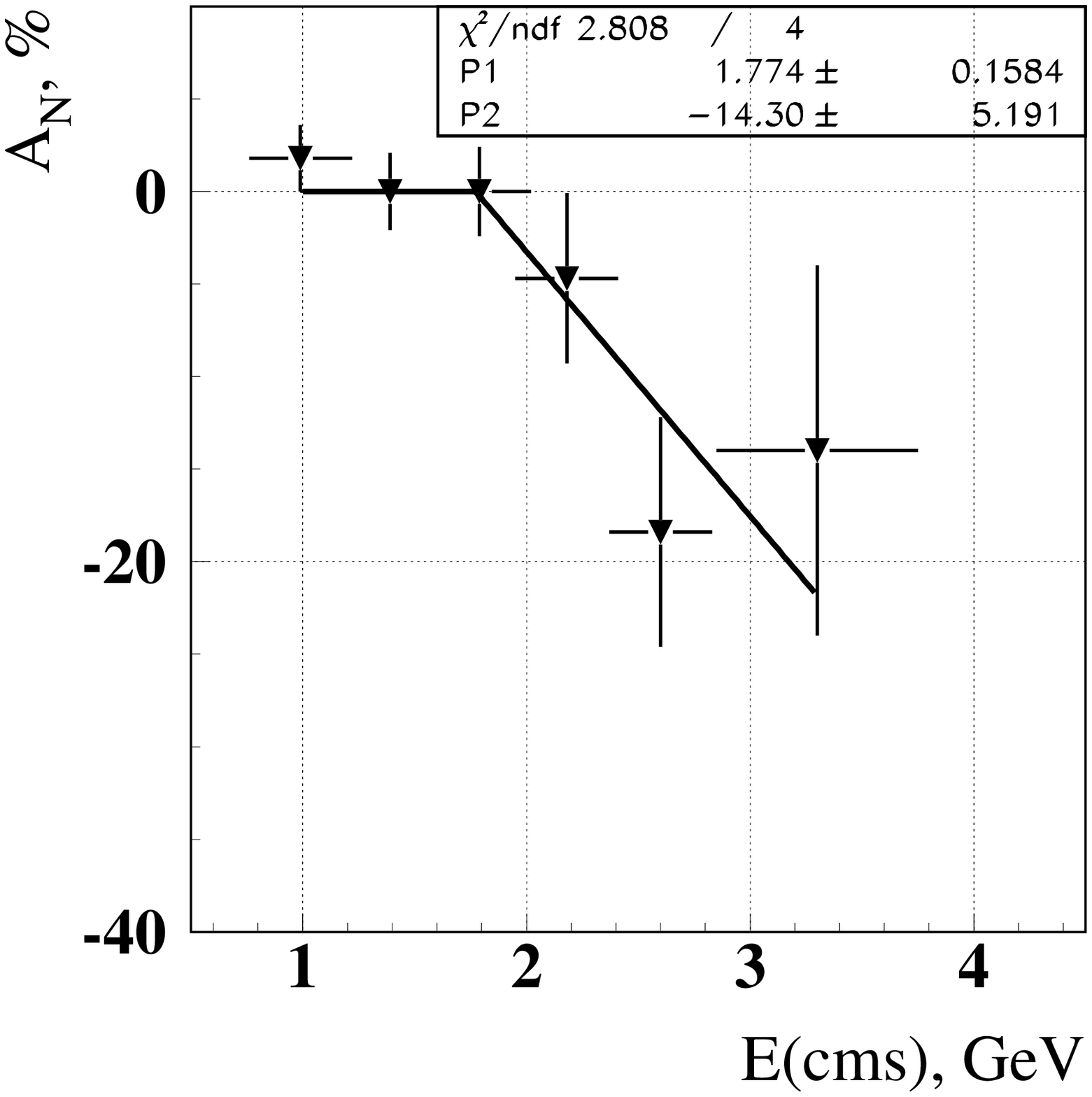} 
\\
\end{tabular}
\caption{The dependence of $A_N$ on  $\pi^0$-meson energy in the c.m.s. 
in the reaction $\pimp$ at the central region  (left) \cite{protv88} 
and in the polarized target fragmentation region (right) \cite{proza40} 
at 40~GeV.}
\label{fig:proza40}
\end{figure}

The result of this fit is presented in {\bf Fig.~\ref{fig:proza40}}.
The asymmetry starts to rise up at $E^0_{cms} = (1.67 \pm 0.11)$~GeV in 
the central region and $E^0_{cms} = (1.76 \pm 0.16)$~GeV in the
target fragmentation region. Really the error bars are higher due to
energy resolution and averaging over transverse momenta and  $\xf$.
For example,  if $\xf$ changes by 0.01, this would bring  0.1~GeV 
dispalcement in the final result at 200~GeV. 

\subsection*{Asymmetry in $\pi^+$ production.}

The result of these two measurements shows that SSA starts to
grow up at the same energy. However  we can not make a 
final conclusion whether the SSA behaviour 
depends on a beam energy or not. We have analysed other experimental data 
to study this threshold effect.
 
A comparison of E704 (FNAL) and E925 (BNL) experimental results is 
presented in {\bf Fig.~2}. The $\pi^+$ asymmetry in the E925 
experiment (22~GeV, \cite{e925}) 
and in  the E704 experiment (200~GeV, \cite{e704fragm})    
starts to rise up at different values of $\xf$ 
($\xf^0 \approx 0.18$ for E704 and $\xf^0 \approx 0.46$ for E925),
but  at the same energy in the centre 
of mass system, $E_0^{cms} \approx 1.6$~GeV. It happened to be 
surprisingly the same energy as for the PROZA-M experiment.    

\begin{figure}[t]
\centering
\begin{tabular}{cc}
\includegraphics[width=0.38\textwidth]
{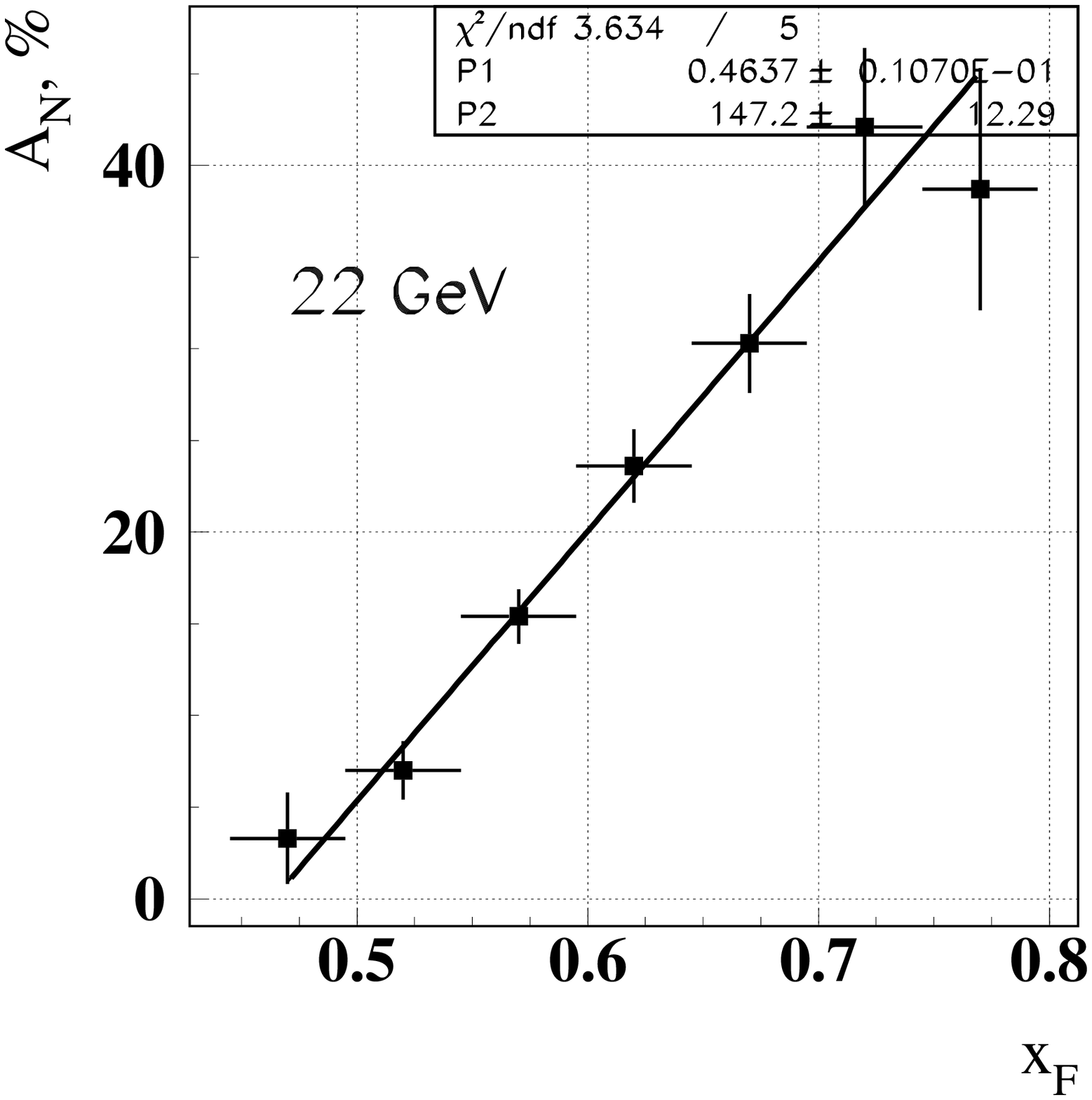} &
\includegraphics[width=0.38\textwidth]
{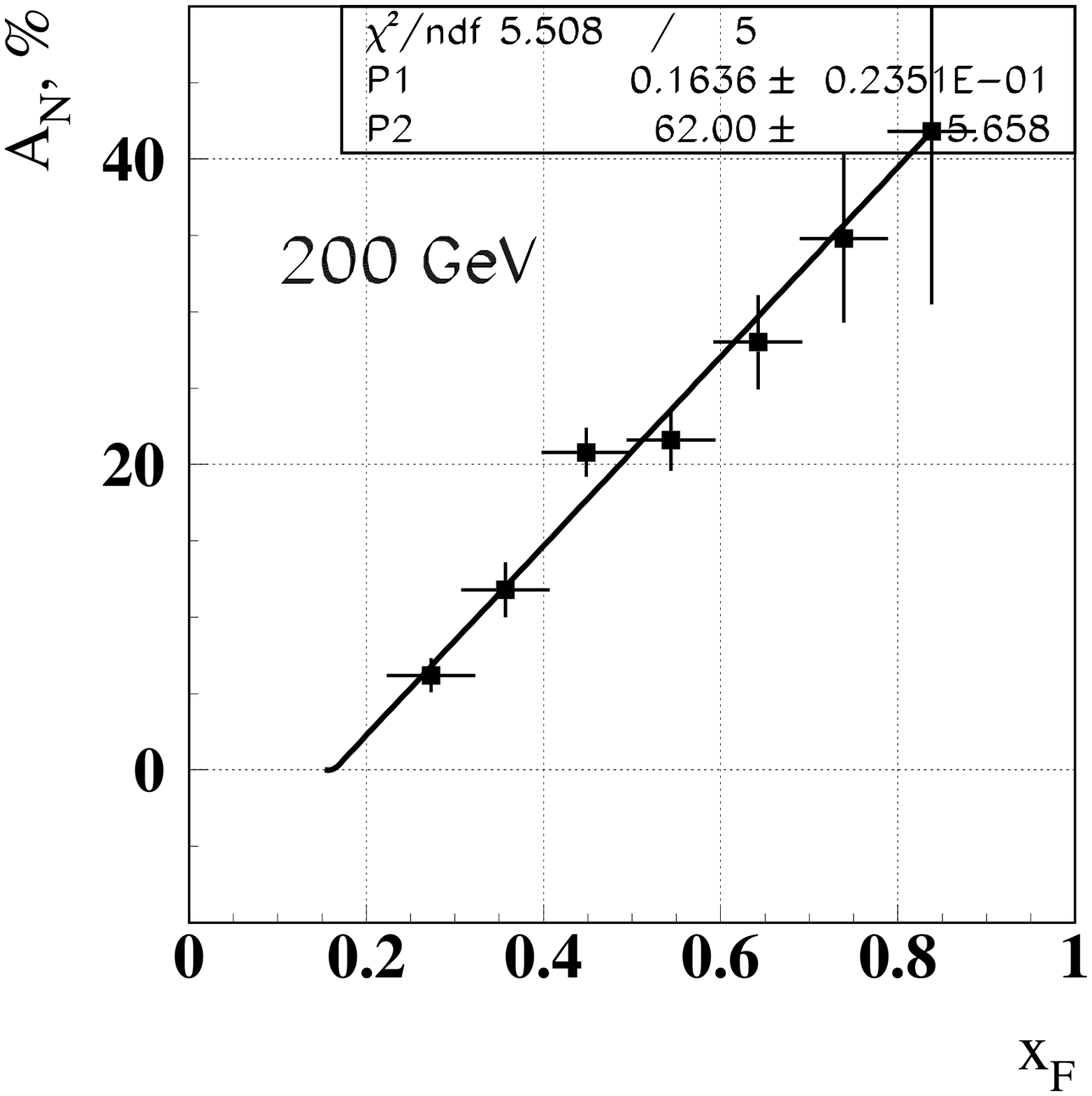}\\
\includegraphics[width=0.38\textwidth]
{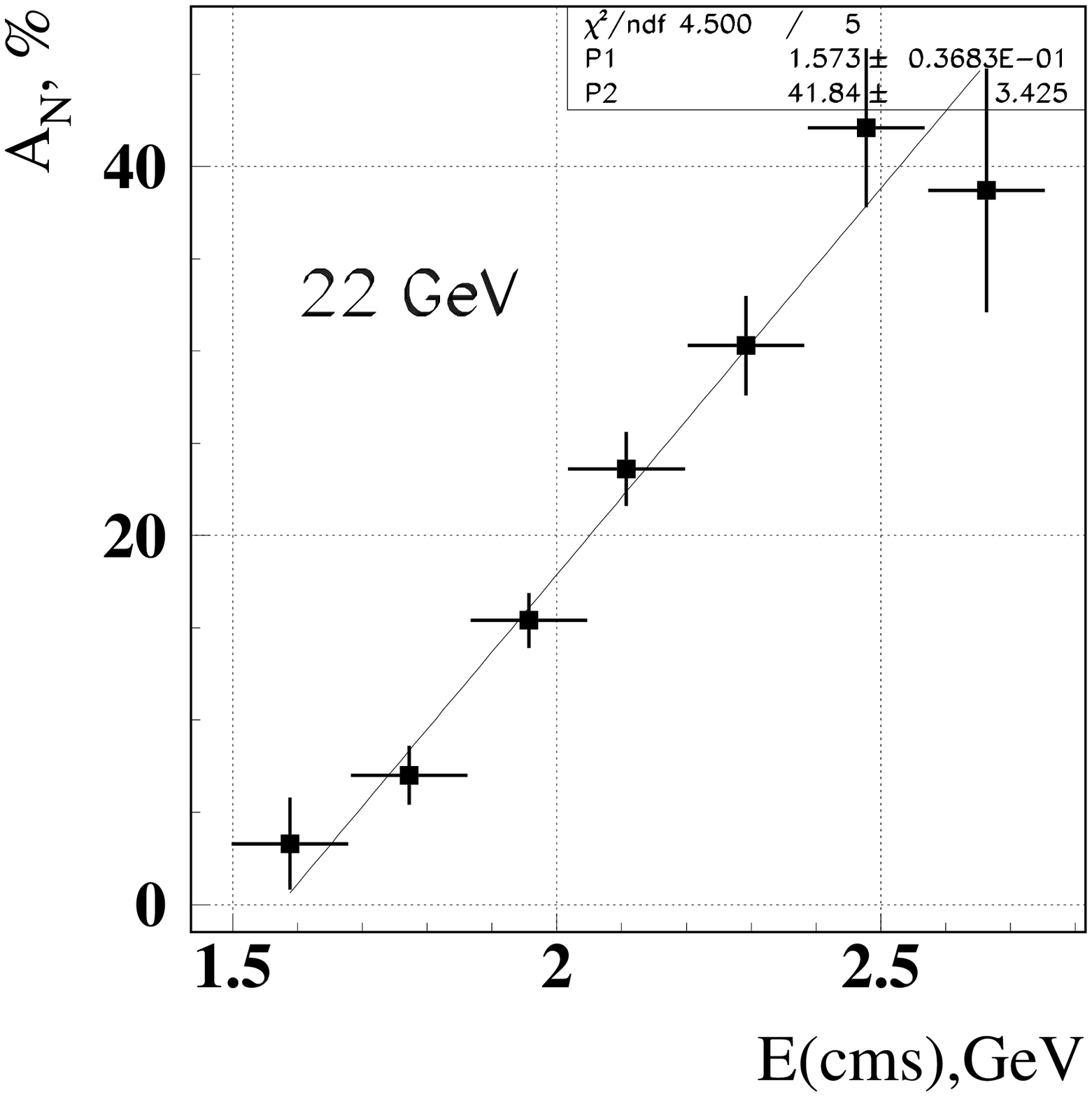} &
\includegraphics[width=0.38\textwidth]
{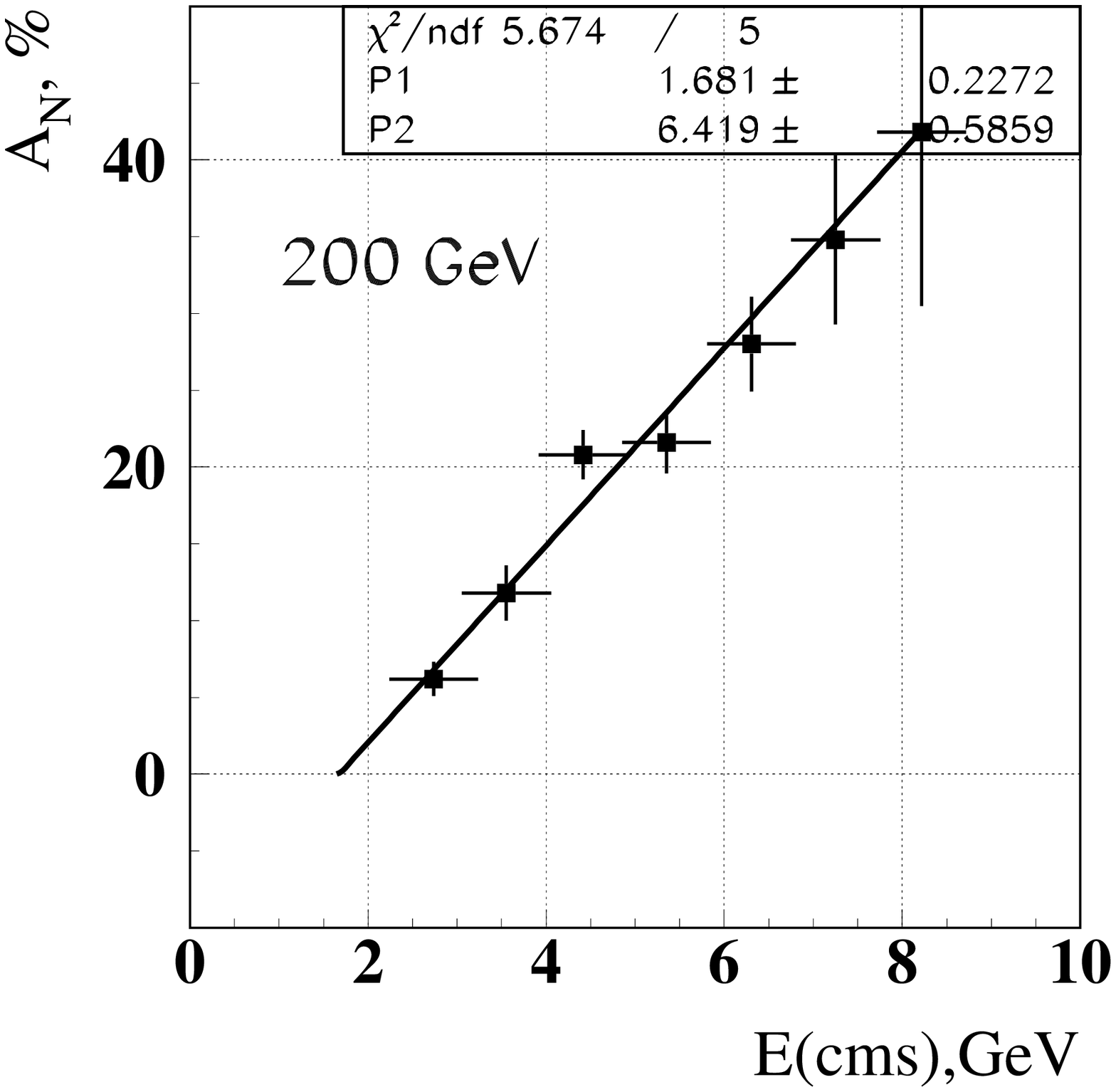}\\
\end{tabular}
\caption{The dependence of  $A_N$ on $\xf$ (top) and energy in c.m.s.
in the reaction  $\pdup + p \rar \pi^+ +X$ in the polarized 
beam fragmentation region from the experiments  E925 at 22~GeV (left) 
\cite{e925} and E704 at 200~GeV (right) \cite{e704fragm}}
\label{fig:xf}
\end{figure}

We used the results of the all fixed target experiments in the energy range 
between 13 and 200~GeV and fit the asymmetry by the function 
(\ref{eq:fit}) regardless that 
in the original papers the asymmetry was presented as a function of
$p_T$ or $\xf$.

\begin{figure}[ht]
\centering
\begin{tabular}{cc} 
\includegraphics[width=0.34\textwidth]
{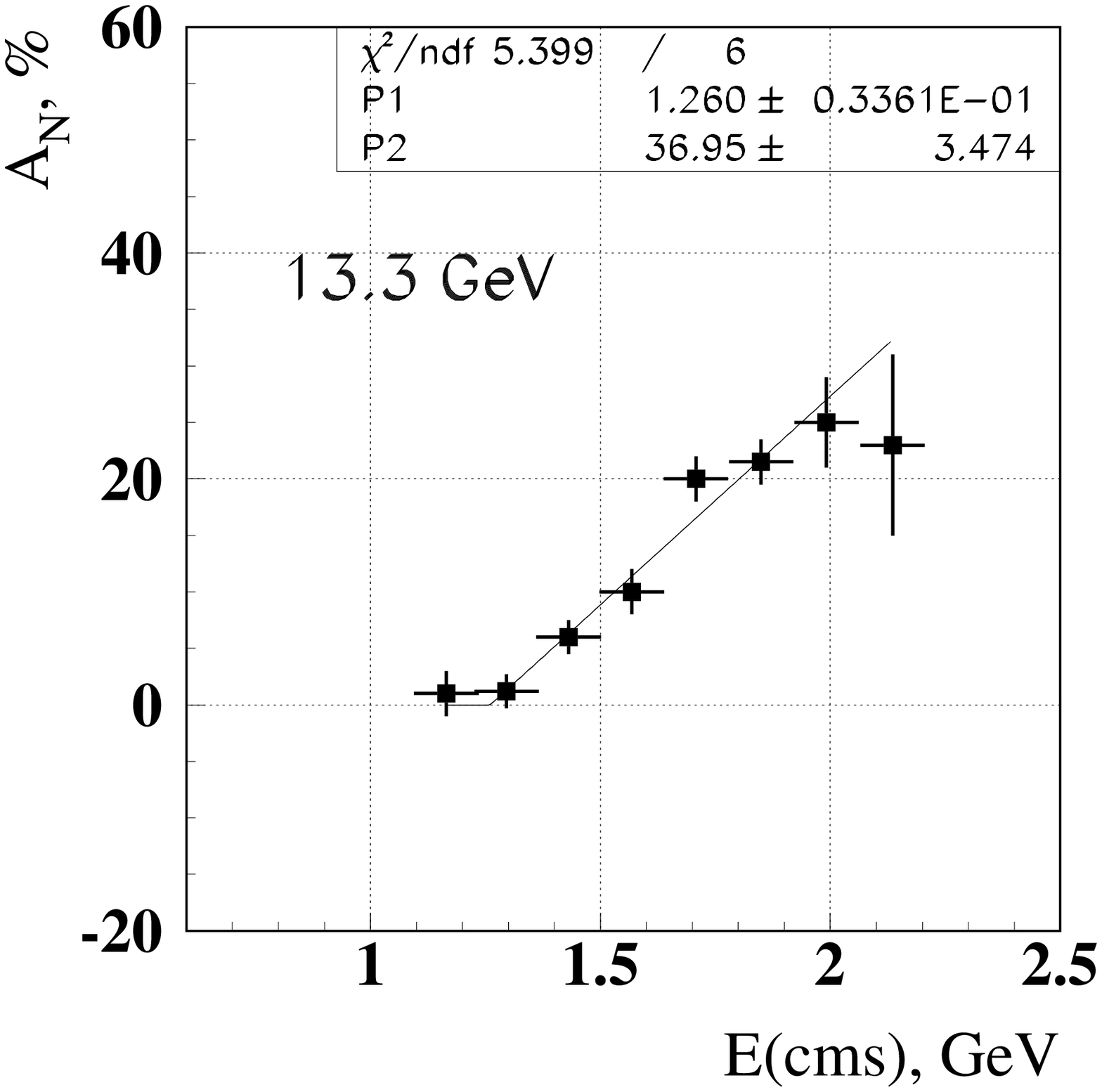} &
\includegraphics[width=0.34\textwidth]
{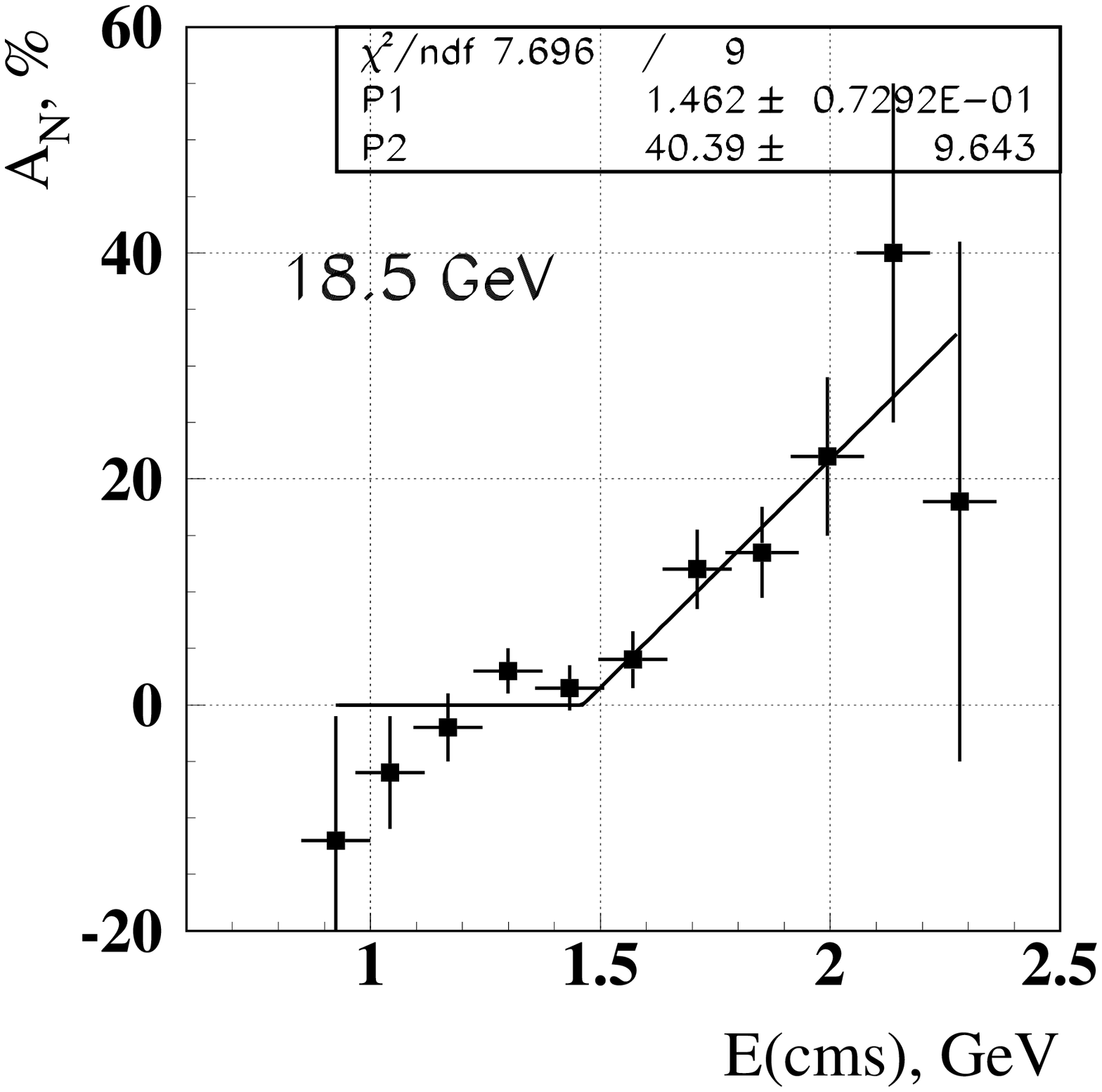} \\
\end{tabular}
\caption{$A_N$ in the reaction 
$\pdup + p \rar \pi^+ + X$ at  13.3 (left) and 
at 18.8~GeV (right) \cite{bnl18}. }
\label{fig:bnl18}
\end{figure}

The dependence of the $\pi^+$ inclusive production asymmetry
on transverse momentum was also studied at BNL at $<\xf>=0.2$
at 13.3 and 18.5~GeV \cite{bnl18} and by the FODS collaboration at Protvino
in the central region \cite{fods}. The asymmetry starts to
rise up at $E_0^{cms} =(1.26\pm0.04)$~GeV at the beam energy of 13.3~GeV 
and at $E_0^{cms}=(1.46\pm0.08)$~GeV at 18.5~GeV 
(see {\bf Fig.~\ref{fig:bnl18}}).

The authors  \cite{fods} claimed that SSA started to rise up 
at $x_T=p_T/\sqrt s =0.37 \pm 0.02$. This value corresponds to 
$(1.62\pm 0.1)$~GeV at $\xf\equiv 0$. Taking into account the  
average  $\xf$ value for each interval we can obtain  
$E_0^{cms}=(1.64\pm 0.15)$~GeV.

\subsection*{Asymmetry in $\pi^0$ production.}

The $\pi^0$ inclusive asymmetry was measured in the $p\pdup$-interaction at 
CERN at 24~GeV \cite{dick24}, in the reaction $\rpip \rar \npionx$ 
at 40~GeV \cite{proza40}, 
in the reaction $p\pdup \rar \npionx$ at 70~GeV \cite{prelim70}  at
Protvino and in the $\pdup p$ and $\bar{p}_{\uparrow} p$ 
interactions at 200~GeV at Fermilab \cite{e704pi0}.

\begin{figure}[h]
\centering
\begin{tabular}{cc}
\includegraphics[width=0.34\textwidth]
{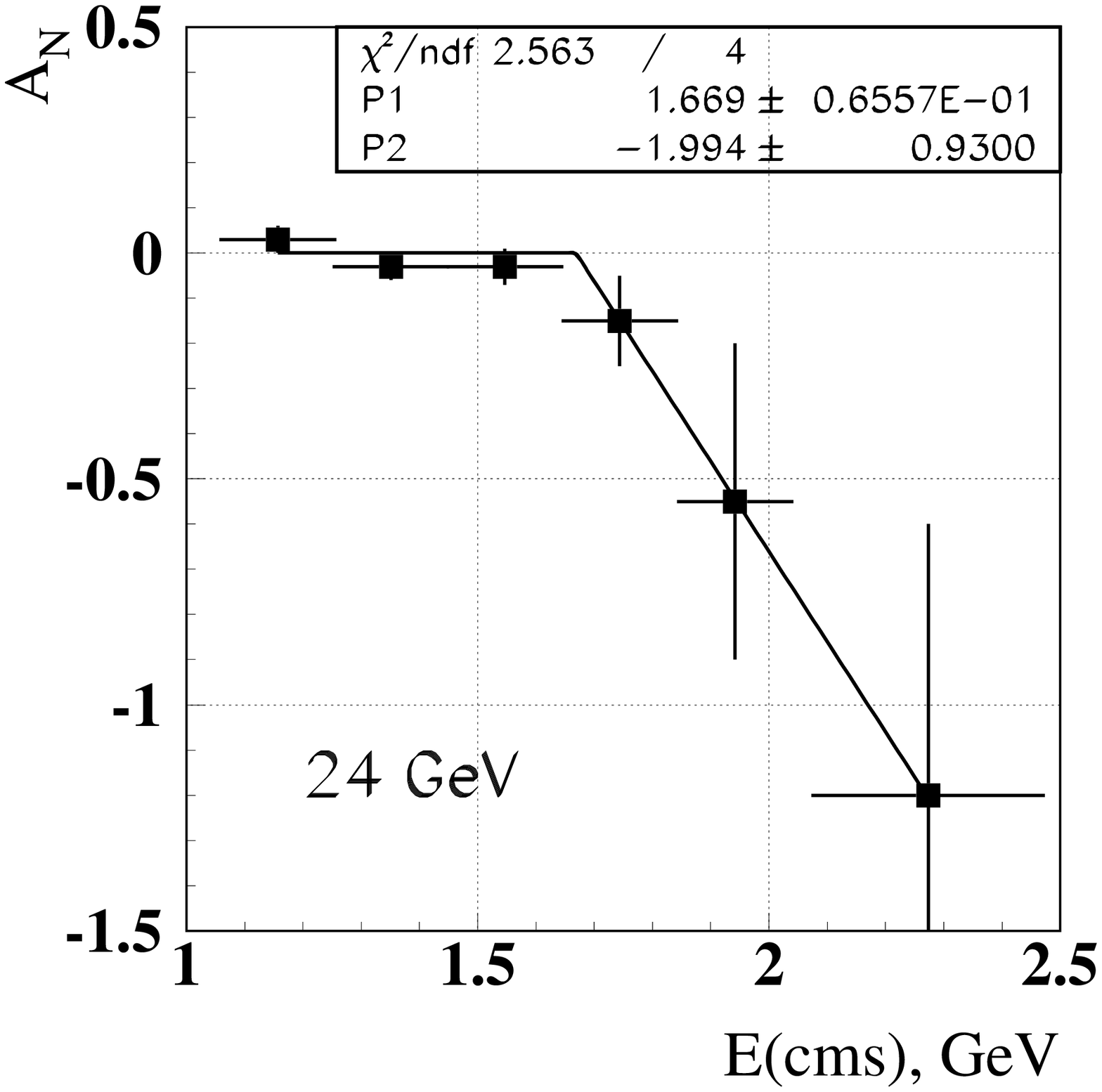} &
\includegraphics[width=0.34\textwidth]
{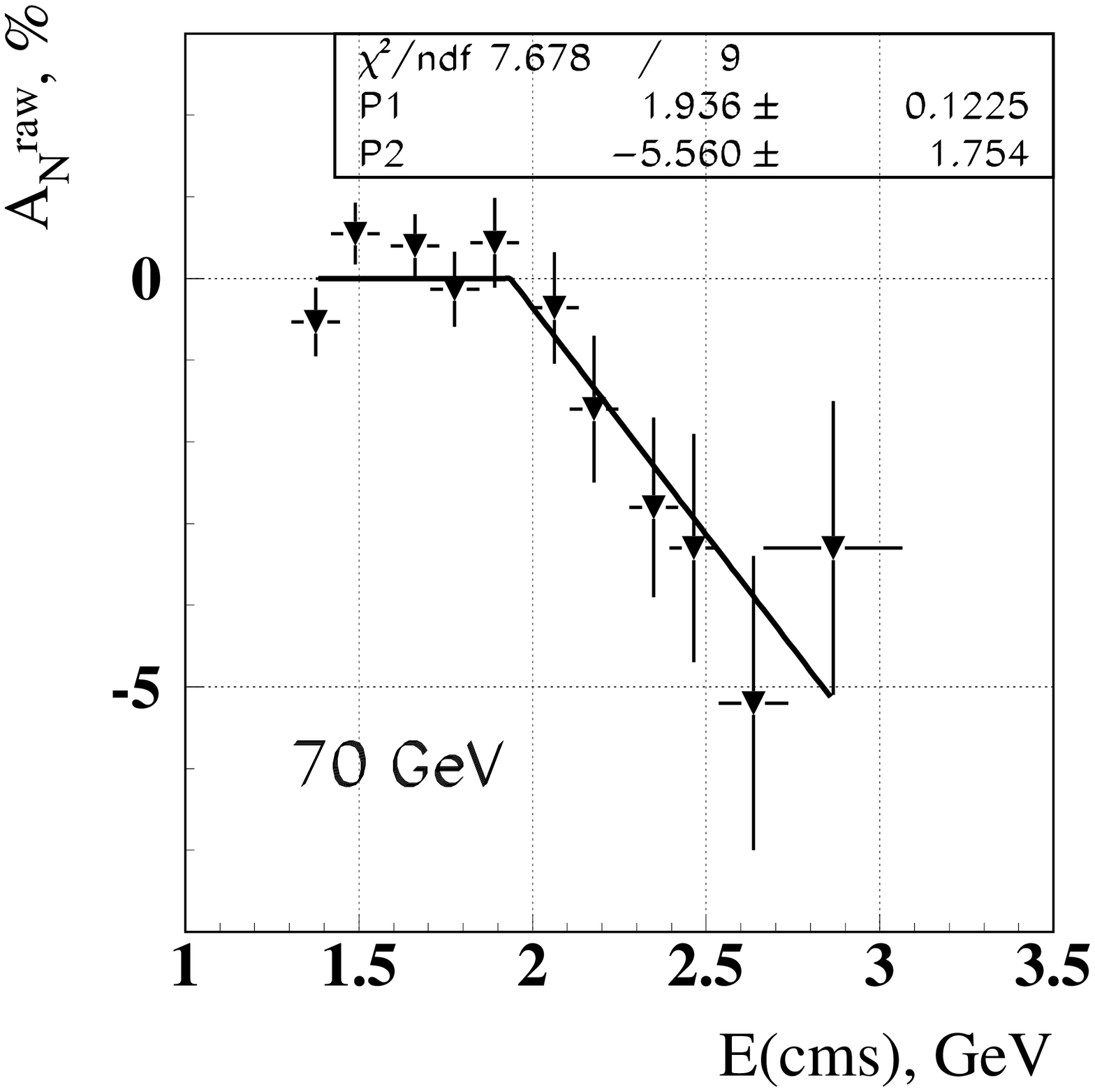} \\
\end{tabular} 
\caption{$A_N$ in the reaction 
$p \pdup \rar \npionx$ in the central region 
at 24~GeV (CERN \cite{dick24}, left) and in the same reaction
in the polarized target fragmentation region at 
70~GeV (Protvino \cite{prelim70}, right).}
\label{fig:pi0other}
\end{figure}

The asymmetry in the reaction $p+\pdup \rar \npionx$ at 24~GeV 
starts to rise up
at  $E_0^{cms}=(1.70 \pm 0.07)$~GeV/c  and at $E^{cms}_0=(1.93 \pm 0.12)$~GeV
at 70~GeV ({\bf Fig. \ref{fig:pi0other}}).

\begin{figure}[t]
\centering
\begin{tabular}{cc}
\includegraphics[width=0.3\textwidth]
{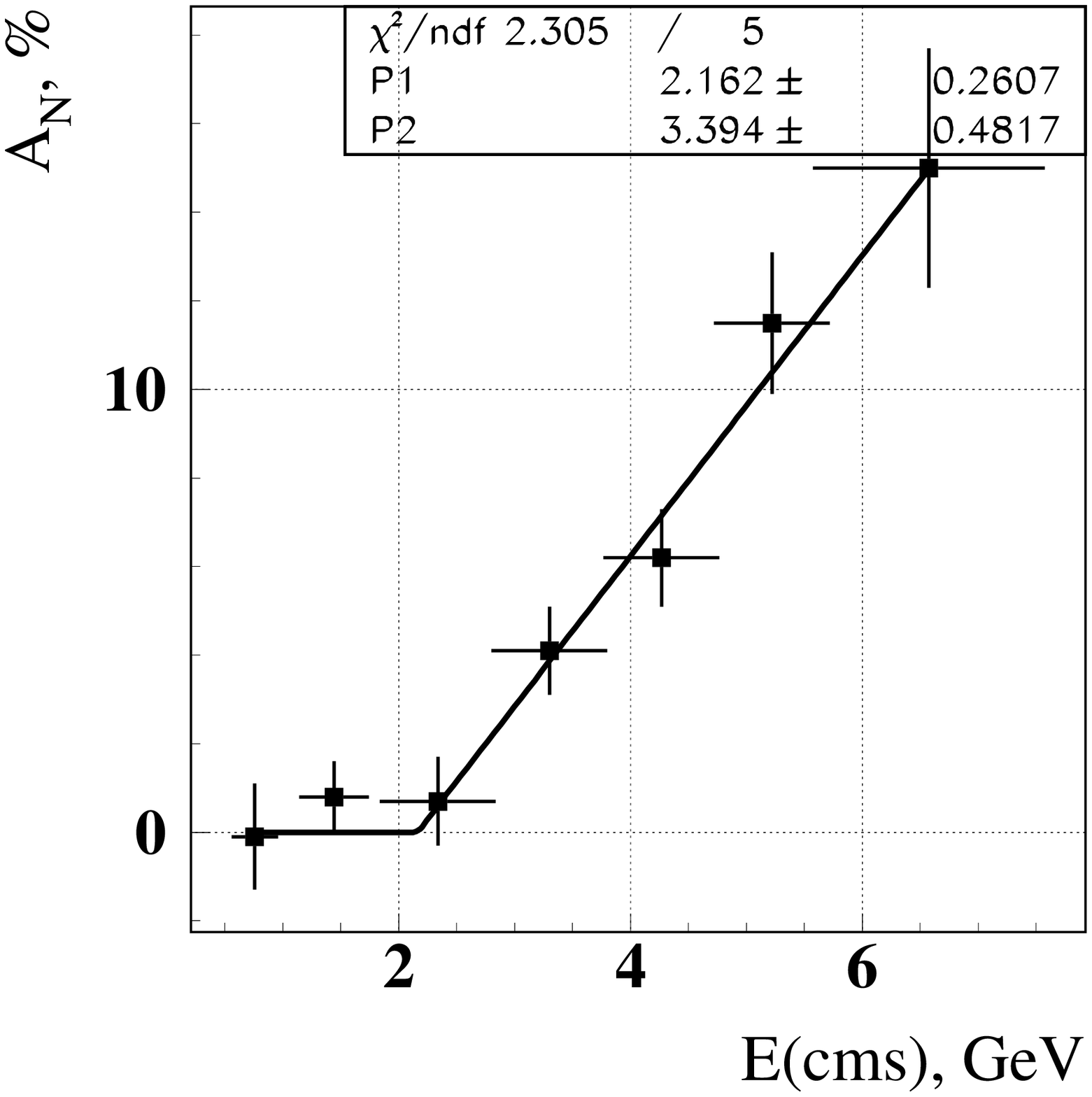} &
\includegraphics[width=0.3\textwidth]
{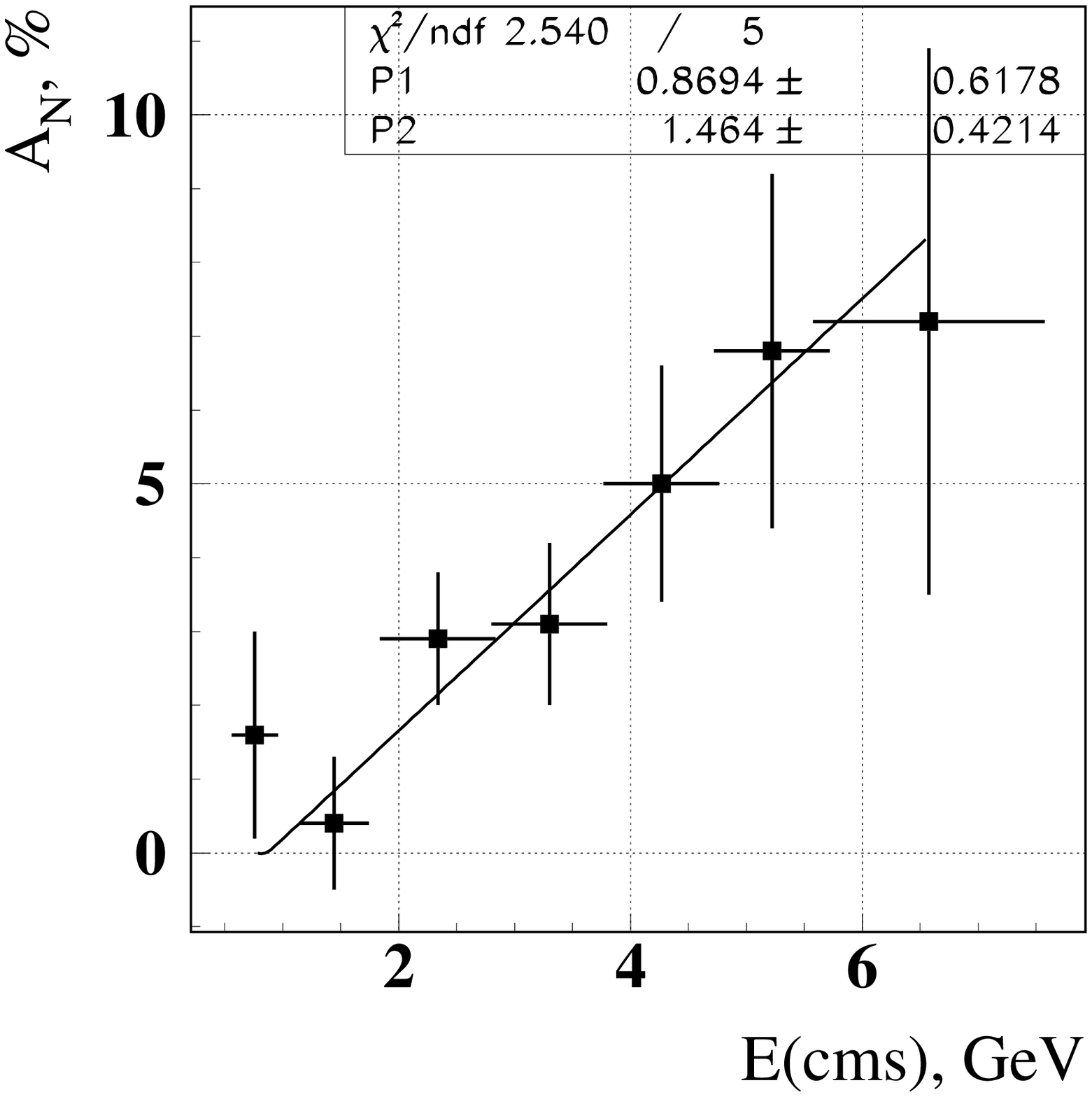}\\
\end{tabular}
\caption{$A_N^{\pi^0}$ in
$\pdup p$ (left) and $\bar{p}_{\uparrow} p$ (right) 
interactions at 200~GeV}
\label{fig:ppanti}
\end{figure}

$A_N$ in the reaction  $\pdup+p \rar \npionx$ 
at 200~GeV \cite{e704pi0}  starts to rise up at 
$E_0^{cms} = (2.16 \pm 0.26)$~GeV, and at
$E_0^{cms} = (0.9 \pm 0.6)$~GeV in the reaction 
$\bar{p}_{\uparrow}+p \rar \npionx$ in the polarized beam 
fragmentation region ({\bf Fig.~\ref{fig:ppanti}}).

The asymmetry in the reaction  $\pdup+p \rar \npionx$ in the central 
region was found to be zero at 70 \cite{proza70} and 200~GeV
 \cite{e704cent}.
 
\subsection*{Asymmetry in $\pi^-$ production.}

\begin{figure}[h]
\centering
\begin{tabular}{cc}
\includegraphics[width=0.35\textwidth]
{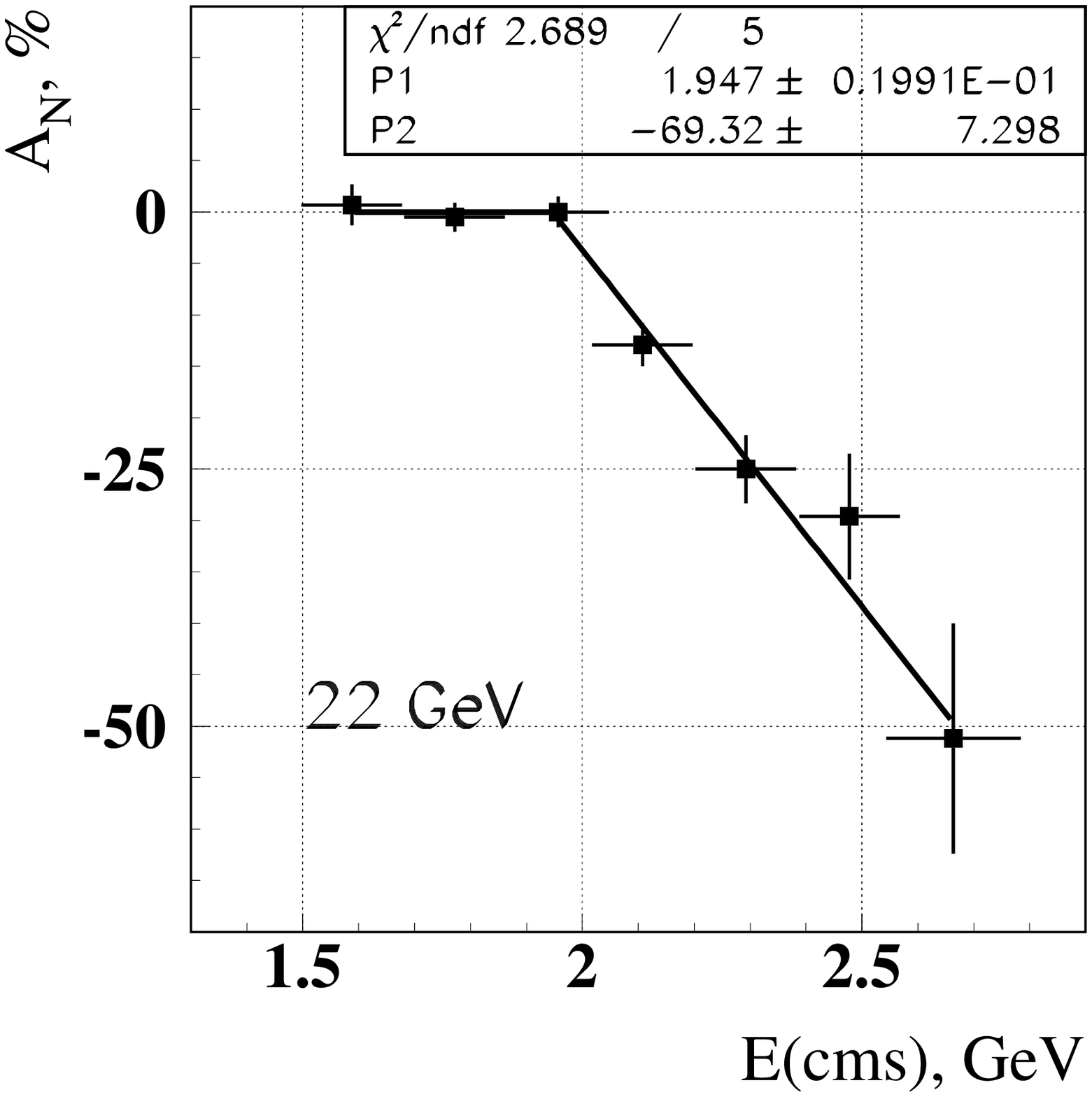} &
\includegraphics[width=0.35\textwidth]
{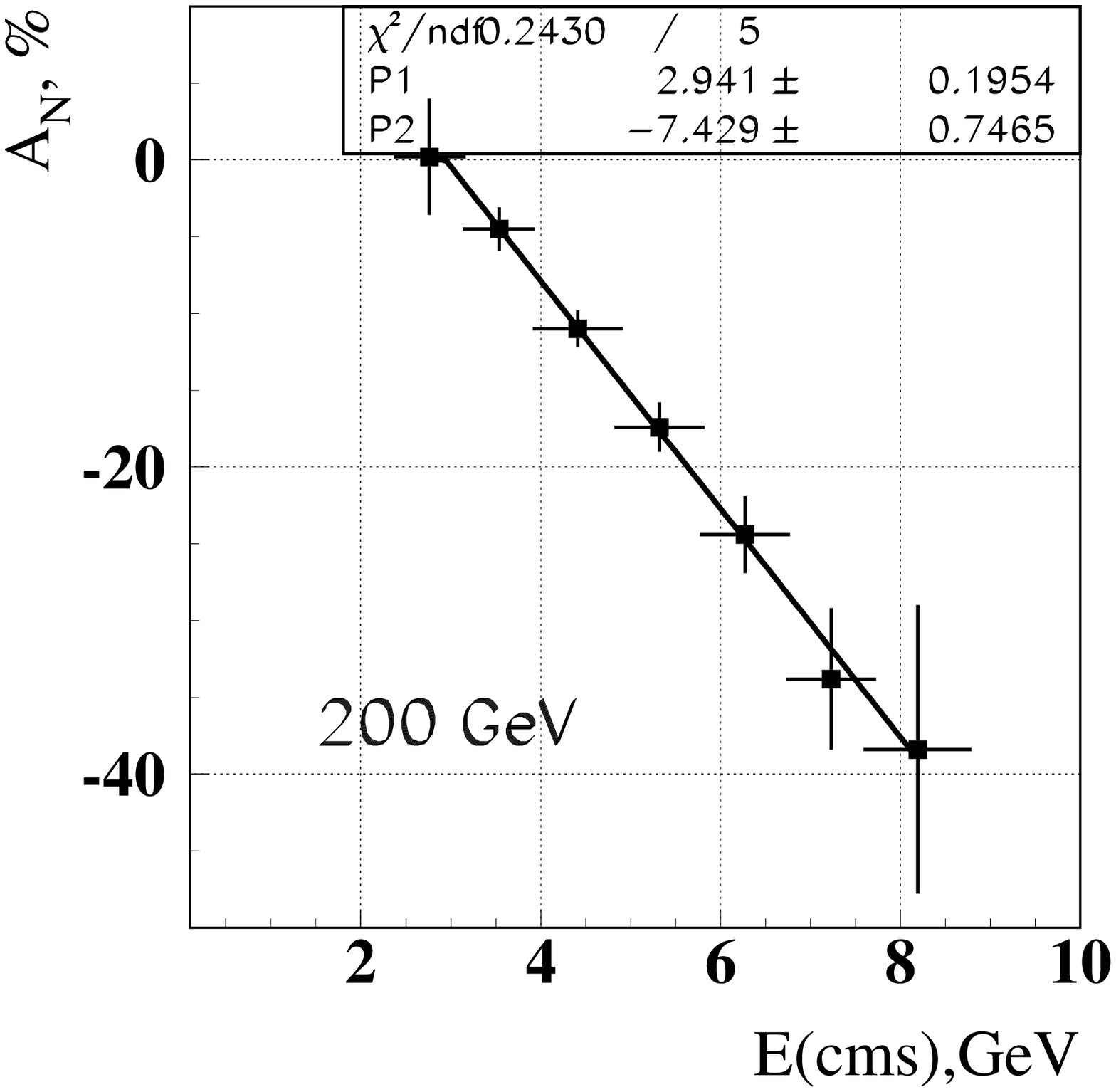}\\
\end{tabular}
\caption{$A_N $ in the reaction  
$\pdup + p \rar \pi^-+X$ in the polarized beam 
fragmentation region at 22~GeV (left, \cite{e925}) and 
200~GeV(right, \cite{e704fragm}).}
\label{fig:piminus}
\end{figure}

The SSA in the reaction  $\pminp$ in the polarized beam fragmentation
region starts to grow up at $E_0^{cms}=(1.95 \pm 0.02)$ at 
22~Gev \cite{e925} and at $E_0^{cms}=(2.9 \pm 0.2)$ 
at 200~GeV \cite{e704fragm} ({\bf Fig.~\ref{fig:piminus}}).

The $\pi^-$ inclusive  asymmetry in this reaction in the central
region was found to be zero at the experiments carried out 
at BNL  at 13.3 and 18.5~GeV~\cite{bnl18}
and at Protvino at 40~Gev \cite{fods}.

\subsection*{Asymmetry in the reaction 
$\bar{p}_{\uparrow} p \rar \pi^{\pm}+X$ at 200~GeV}

\begin{figure}[ht]
\centering
\begin{tabular}{cc}
\includegraphics[width=0.35\textwidth]
{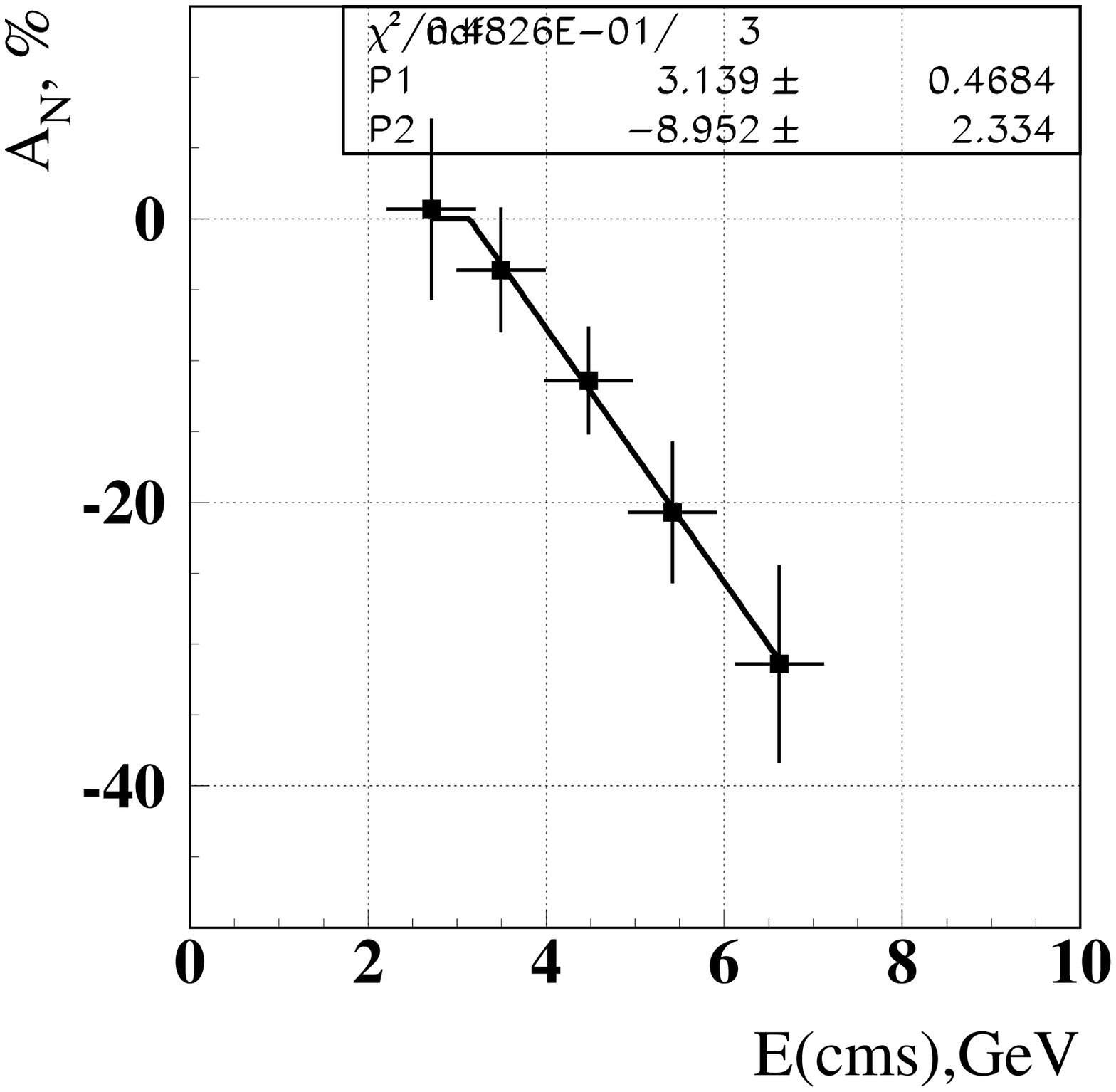} &
\includegraphics[width=0.35\textwidth]
{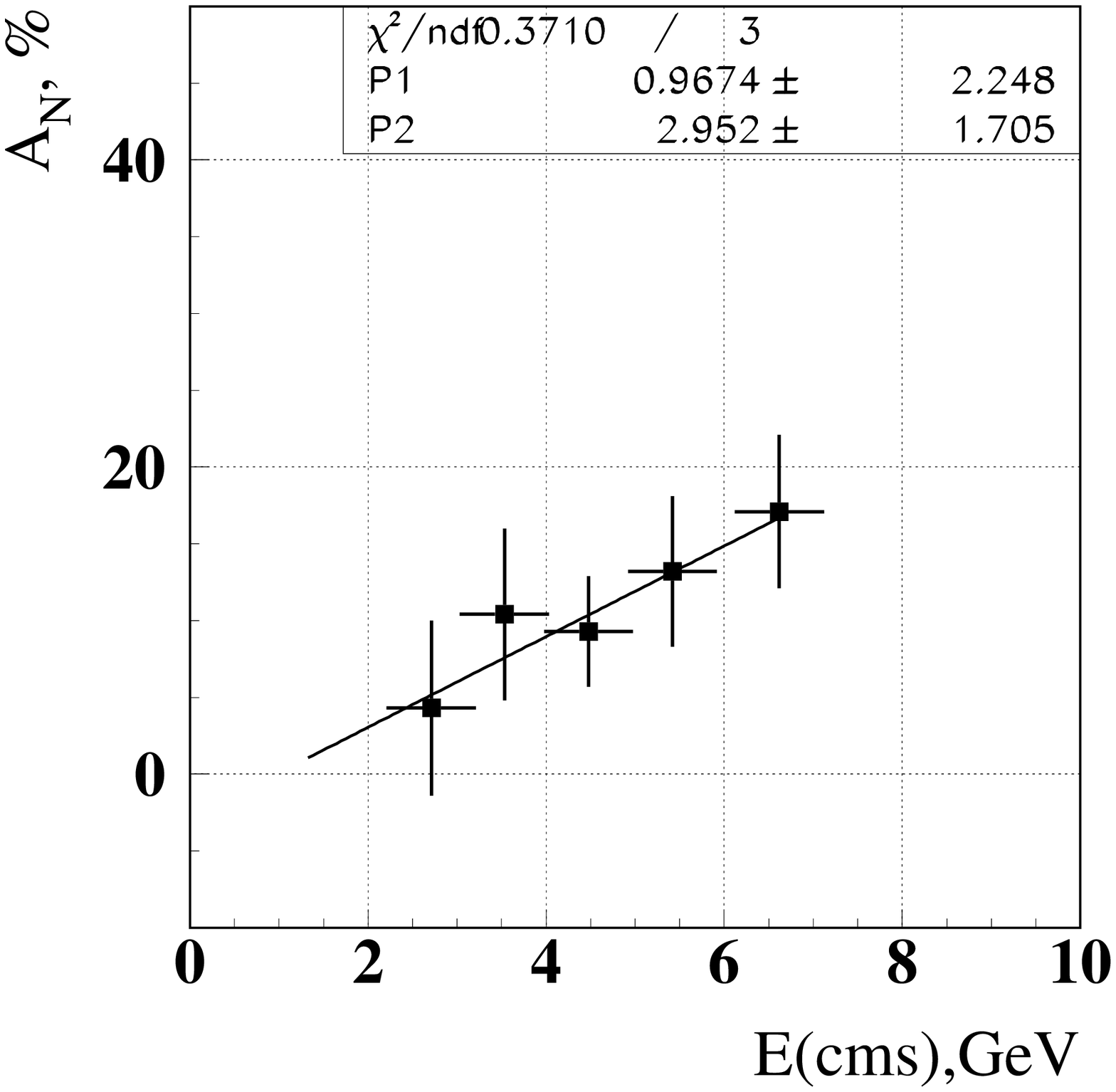} \\
\end{tabular}
\caption {$A_N^{\pi^+}$ (left) and  $A_N^{\pi^-}$ (right) 
in the $\bar{p}_{\uparrow}p$ interaction in the beam fragmentation
region at 200~GeV \cite{e704_anti}.}
\label{fig:e704plus_anti}
\end{figure}

The SSA in the reaction  $\bar{p}_{\uparrow} p \rar \pi^{\pm}X$ 
in the beam fragmentation region was measured at Fermilab
at 200~GeV ({\bf Fig.~\ref{fig:e704plus_anti}}). 
The asymmetry starts to rise up at
$E_0^{cms}=3.1 \pm 0.5$ for $\pi^+$ and 
at $E_0^{cms}=1.0 \pm 2.2$~GeV for $\pi^-$-meson.

\subsection*{Discussion.}

\begin{figure}[htb]
\centering
\includegraphics[width=0.75\textwidth]
{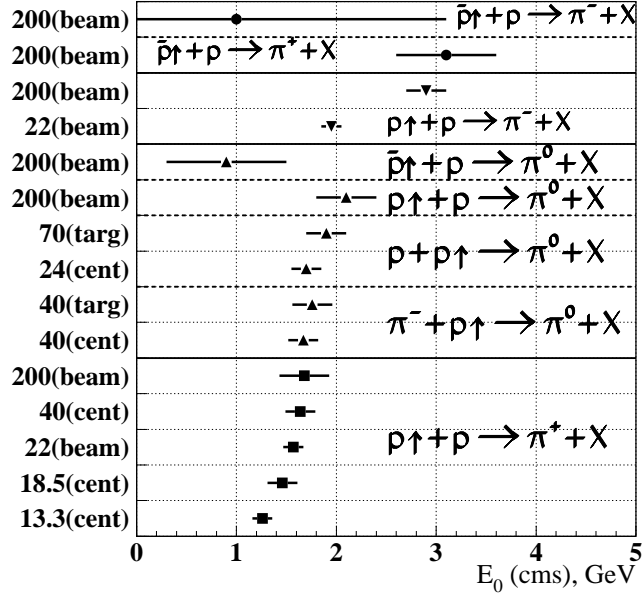} 
\caption {Centre of mass energy values where  the pion asymmetry 
begins to grow up for different experiments.
The energy along the Y-axis is in GeV; $cent$ -- corresponds to 
experiments in the central
region ( $x_f\approx0$), $targ$ -- the polarized target
fragmentation region; $beam$ -- the polarized beam fragmentation region. }
\label{fig:summary}
\end{figure}

The combined result of our analysis is presented in  
{\bf Fig.~\ref{fig:summary}} and {\bf Table~\ref{tab:summary}}.
The error bars include both the fit procedure errors and
the resolution of kinematic parameters. The values of
$\chi^2/N$ and the slope of $k \cdot (\sqrt s -E_{cms}^0)$ (asymptotic asymmetry
at the phase space limit whithout saturation effect) are also presented
in the table. We did not include the experiments where 
the asymmetry is close to zero in this table.

The main conclusion is that the asymmetry starts 
to grow up at the same centre of mass energy $E_0^{cms}=1.5$ to $2.0$~GeV for 
the most of the experiments in the energy range between 13 and 200~GeV. 
The analysis was done only for those experimental data where a transverse 
momentum $p_T$ was greater than 0.5~GeV/c to exclude soft interactions. 
The conclusion is valid for all $\pi^+$ and $\pi^0$ asymmetries. 
We have to mention that $\pi^-$ production seems to contradict to this. 
We can explain this fact that $\pi^-$-meson at 
small $x_F$ can be produced not only from the valence $d$-quark but also from 
other channels. The interference of different channels is also responsible for 
asymmetry cancellation in $\pi^0$ and $\pi^-$ production in the 
central region. In the reaction $\pimp$ in the central region we found 
significant asymmetry in the contrary to the $\pdupp$ reaction. If in  
the $\pdupp$ reaction the asymmetry is 
cancelled because of different channel interference 
from a polarized and non-polarized proton, in the $\pi^-\pdup$ collisions
the valence $u$-quark from a polarized proton combining with the valence 
$\bar{u}$-quark from $\pi^-$ gives the main contribution to $\pi^0$ 
production, while other channels are suppressed.

\begin{table}[t]
\centering{
\caption{Summary table. 
Centre of mass energy values $E_{cms}^0$ where  the pion asymmetry 
begins to grow up for different experiments. 
$E^{max}_{cms}=\sqrt s/2$.  
}
\label{tab:summary}
\begin{tabular}{|c|c|c|c|c|c|}
\hline
\hline
Reaction & Energy & 
$E_{cms}^0$, GeV & $\chi^2 /N$ 
& $k \cdot (E^{max}_{cms} - E^0_{cms})$, \%
&Ref.\\
\hline
\hline
$\pdup +p \rar \pi^+ + X$ & 13.3 & 
$ 1.26\pm0.1$ & 0.9 & $52 \pm 6$ &
\cite{bnl18}\\
$\pdup +p \rar \pi^+ + X$ & 18.5 & 
$ 1.46\pm0.15$ & 0.85 & $63 \pm 16$ &
\cite{bnl18}\\
$\pdup +p \rar \pi^+ + X$ & 21.92& 
$ 1.57 \pm 0.1$ & 0.9 & $68 \pm 6$ &
\cite{e925}\\
$\pdup +p \rar \pi^+ + X$ & 40 &   
$ 1.64 \pm 0.15$ & &  &\cite{fods}\\
$\pdup +p \rar \pi^+ + X$ & 200  & 
$ 1.68 \pm 0.25$ & 1.1 & $52 \pm 5$ &
\cite{e704fragm}\\
\hline
$\pi^- +\pdup \rar \pi^0 + X$ &40 & 
$ 1.67 \pm 0.15$ & 1.5 & $107 \pm 26$ &
\cite{protv88}\\
$\pi^- +\pdup \rar \pi^0 + X$ &40 & 
$ 1.76 \pm 0.2$ & 0.7 & $36 \pm 14$ &
\cite{proza40}\\
$p +\pdup \rar \pi^0 + X$ &24 & 
$ 1.7 \pm 0.15$ & 0.6 & $334 \pm 165$ &
\cite{dick24}\\
$p +\pdup \rar \pi^0 + X$ &70& 
$ 1.9 \pm 0.2$ & 0.85 & $208 \pm 70$ &
\cite{prelim70}\\
$\pdup+p \rar \pi^0 + X$ &200 & 
$ 2.1 \pm 0.3$ & 0.5 & $26 \pm 5$ &
\cite{e704pi0}\\
$\bar{p}_{\uparrow}+p  \rar \pi^0 + X$ &200 &
$ 0.9 \pm 0.6$ & 0.5 & $13 \pm 4$ &
\cite{e704pi0}\\
\hline
$\pdup +p \rar \pi^- + X$ & 21.92& 
$ 1.95 \pm 0.1$ & 0.5 & $87 \pm 11$ &
\cite{e925}\\
$\pdup +p \rar \pi^- + X$ & 200  & 
$ 2.9 \pm 0.2$ & $<$0.1 & $51 \pm 6$ &
\cite{e704fragm}\\
\hline
$\bar{p}_{\uparrow}+p \rar \pi^+ + X$ &200 & 
$ 3.1 \pm 0.5$ & $<$0.1 & $59 \pm 16$ &
\cite{e704_anti}\\
$\bar{p}_{\uparrow}+p \rar \pi^- + X$ &200 & 
$ 1.0 \pm 2.2$ & 0.1 & $25 \pm 15$ &
\cite{e704_anti}\\
\hline
\hline
\end{tabular}
}
\end{table}

In  this scheme the asymmetry behaviour in ${\bar{p}_{\uparrow}p}$ 
interactions in $\pi^+$ and $\pi^-$ production should be inversed in 
comparison with the $\pdupp$ data. The result from E704 
experiment \cite{e704_anti} is consistent with this model. 
The asymmetry of $\pi^+$-production starts to grow up at the same value 
$E_0^{cms} \approx 2.9$~GeV as for $\pi^-$ in reaction $\pdupp$, and 
the asymmetry in the reaction ${\bar{p}_{\uparrow}+p \rar \pi^- +X}$ begins 
to rise up at small value $E^0_{cms}$. 

We can conclude that the meson asymmetry produced by valence 
quark starts to grow up at the same universal energy $E^0_{cms}$.  
Also the values of the parameter $k \cdot (E^{max}_{cms} - E^0_{cms})$ 
are close  for all eight measurements in the reactions $\pplup$ and
$\pminp$.

We are thankful to Yu.~Matulenko, L.~Nogach, P.~Semenov,
L.~Soloviev, K.~Shestermanov and A.~Zaitsev for fruitful discussions.

{\it The work is supported by Russian Foundation for 
Basic Research grant 03-02-16919}.



\begin{thebibliography}{99}
\bibitem{protv88}~V.D.~Apokin et al., \PLB {\bf 243}, 461 (1990).
\bibitem{proza40}~A.N.~Vasiliev et al.,  IHEP Preprint 2003-21, 
Protvino, 2003; accepted to Sov. Journ. Nucl. Phys. (in Russian);\\
V.~Mochalov at al, presented on the X Workshop on High Energy
Spin Physics, Dubna, September 2003.
\bibitem{prelim70} ~N.~Belikov et al, preprint IHEP-1997-51, Protvino, 1997;
~N.~Belikov et al., Proc. of  13th International Symposium on 
High-Energy Spin Physics (SPIN 98), Protvino, Russia, 8-12 Sep 1998.
In *Protvino 1998, High energy spin physics* 465-467.
\bibitem{e925} ~C.E.~Allgower et al., \PRD {65}(2002),092008.
\bibitem{e704fragm}~D.L.~Adams et al., \PLB {264} (1991), 462. 
\bibitem{bnl18}~S.~Heppelmann et al., SINGLE SPIN ASYMMETRY IN LARGE 
P(T) INCLUSIVE PI+ AND PI- PRODUCTION FROM P (POLARIZED) P INTERACTIONS, 
Prepared for 8th International Symposium on High-energy Spin Physics, 
Minneapolis, MN, 12-17 Sep 1988. In *Minneapolis 1988, Proceedings, 
High-energy spin physics, vol. 1*, pp. 157-159. 
\bibitem{fods}~V.V.~Abramov et al., Preprint IHEP-96-82, Protvino, 1996; 
Nucl.Phys. B {\bf492} (1997), 3; {\mbox{\textsf{hep-ex/0110011}}}. 
\bibitem{dick24}~J.~Antille et al., \PLB {94} (1980),523.
\bibitem{e704pi0}~D.L.~Adams et al., FERMILAB-PUB-91-014-E; 
ANL-HEP-PR-91-09; IFVE-91-50; \PLB {276}(1992), 531. 
\bibitem{proza70}~A.N.~Vasiliev et al.,  IHEP Preprint 2003-22, 
Protvino, 2003; accepted to Sov. Journ. Nucl. Phys. (in Russian);\\
V.~Mochalov at al, presented on the X Workshop on High Energy
Spin Physics, Dubna, September 2003.
\bibitem{e704cent}~D.L.~Adams et al., Preprint IHEP 91-49;
Z.Phys.C {\bf 56}(1992), 181. 
\bibitem{e704_anti}~A.~Bravar et al., \PRD {55} (1997),1159. 
\bibitem{struct_functions} D. Adams et al. (By Spin Muon Collaboration),
CERN-PPE-97-022; CERN-PPE-97-22; DAPNIA-SPHN-97-27;
\PRD {56} (1997), 5330;
{\mbox{\textsf{ hep-ex/9702005}}}. 
\end{thebibliography}
\end{document}